\documentclass{article}
\usepackage[english]{babel}
\usepackage{amssymb}
\usepackage{amsfonts}
\usepackage{amsmath}
\usepackage{latexsym}

\title{Non-trivial class of the mixed \\$U(\sigma{+}\mu)$-vector solitons.}

\author{A.\,M.\ Agalarov \thanks{agalarov@itp.ac.ru} 
\\M.V.Lomonosov Moscow State University, 119899, Moscow, Russia \\
\\
\and R.\,M.\, Magomedmirzaev, 
\\Institute of physics of DSC of RAS, 367025, Makhachkala, Russia}
\date{}

\begin{document}
\maketitle

\abstract{There has been found an exact solution of the mixed
problem for Shr\"{o}dinger's compact $U(m)$-vector nonlinear model
with an arbitrary sign of the coupling constant. It is shown, that
in case of $m{\geqslant}3$ there is a new class of solutions -
mixed $U(\sigma{+}\mu)$-vector solitons with "inelastic" (changing
the form without the energy loss) interaction at $\sigma{>}1$ and
strict elastic - at $\sigma{=}1$. They correspond to the color
complexes consisting of $\sigma$-bright and $\mu$-dark solitons
($\sigma{+}\mu{=}m$) and they can exist both in self-focusing and
defocusing medias. The universal $N$-soliton formula for the
attraction and repulsion cases has been obtained by the method of
Hirota for the first time.}


\indent PACS{42.65.Tg, 42.65.Hw, 03.40.Kf}
\\
\\
\\
\\
\indent 0. The evolutionary system of coupled nonlinear
Shr\"{o}dinger equations (NLS-$m$)


\begin{equation}
i\hat{L}_{j}\psi_{j}^{*}=\sum_{k=1}^{m}\
a_{jk}\vert\psi_k\vert^{2}\psi_{j}^{*},~j=\overline{1,m}
\end{equation}
\begin{displaymath}
\psi_j\in\mathbf{C},~\hat{L}_{j}=\partial_{\zeta}+\emph{i}\
c_j\partial_{\xi\xi};~ a_{jk},c_{j}\in\mathbf{R}
\end{displaymath}
arises within the weak coupling limits in various nonrelativistic
models of the nonlinear field theory.The integrability conditions
and the exact solutions of NLS are of a broad practical interest
(nonlinear optics, plasma, ferromagnetism, hydrodynamics,
Bose-Einstein atomic condensates etc.[1-6]) together with an
academic one. A strict mathematical derivation of the two
connected parabolic motion equations that are equivalent to the
system (1) when $m=2$ and $c_1 = c_2$ is given in [7] where the
self-influence of different polarization waves in the nonlinear
media of a tensor response has been studied for the first time. At
(1) $c_j$ parameters determine the dispersion, and matrix
coefficients $a_{jk}$ when $j \neq k$ and $j=k$, determine
nonlinear interaction and self-action of $\psi_j$ fields
correspondingly. Subject to the implication of the variable
$\zeta$,$\xi$ and the sign of the parameters  product $sign(c_j
a_{jk})=\ae_k$ the system (1) at a classical level describes a
spatial or time evolution of the $m$-component field in the
nonlinear cube-medium[7,8]; at the quantum one - Bose-gas with
$m$-color degrees of freedom [9,10] with attractive ($\ae_k
>0$) or repulsive ($\ae_k <0$) point interaction.

\indent The exact integrable cases of the system (1), by Liouvill
implication, are highly limited and require satisfaction of the
rigid conditions in space of the controlled parameters ${c_j,
a_{jk}}$. On the bases of Zakharov's theorems about the additional
motion invariants one may show (the proof will be given in a
separate work), that when the conditions are met $ a_{jk}=\pm
a_{kk}, c_j=\pm c$ the system (1) admits representation of a zero
curvature (Lax-pair) and is embeddable into the scheme of the
inverse scattering method (ISM). The integrable reductions of
NLS-$m$ arising in this case form the family of solitons vector
models with a unitary $U(m)$ and pseudounitary $U(m,n)$ group of
symmetry. The known exact solutions of the given family of models,
for example, those of the $U(2)$-vector model of Manakov [8] ($L_0
= L_1 = L_2,~\sigma=2,~\mu=0$)


\begin{equation}
i\hat{L}_{0}\psi_{j}^{*}=
\ae(\vert\psi_1\vert^{2}+\vert\psi_2\vert^{2})\psi_{j}^{*},~j=\overline{1,2}
\end{equation}

are "single-color" multisolitons: bright solitons $(\psi_{1,2}\sim
sech{\alpha})$ in a self-focusing medium $(\ae>0)$  [8] and dark
ones $(\psi_{1,2}\sim \tanh{\beta})$ in a defocusing medium
$(\ae<0)$ [11,12]. Bright-vector [8] and dark-vector [11,12]
soliton solutions that are constructed on the base of the
traditional boundary problems of the $U(1)$-scalar bright
$(\psi(\pm\infty)=0)$ and dark
$(\psi(\pm\infty)=\rho\exp(\emph{i}\Theta))$ solitons [5,13] may
be classed as trivial ones in the given sense. The exact solutions
of the mixed vector solitons in the pseudo-Euclidean
$U(1,1)$-model [9] and in the Euclidean $U(1{+}1)$-model (2) with
a defocusing $(\ae<0)$ nonlinearity have demonstrated [12] that
they also interact (as $U(1)$-scalar solitons) in the elastic
(trivial) fashion.

\indent For the first time it was shown in the given work that in
the family of $U(m)$-vector nonlinear models of Shr\"{o}dinger
(integrable reductions of the system (1))


\begin{equation}
i\hat{L}_{0}\psi_{j}^{*}= \ae\sum_{k=1}^m
\vert\psi_k\vert^{2}\psi_{j}^{*},~j=\overline{1,m}
\end{equation}

with the boundary conditions of the mixed density


\begin{equation}
\psi_{\sigma}(\zeta,\xi)\bigg\vert_{\vert\xi\vert\rightarrow
\infty}{\longrightarrow} 0, ~~~~
\psi_{\mu}(\zeta,\xi)\bigg\vert_{\vert\xi\vert\rightarrow
\infty}{\longrightarrow}\rho_{\mu}e^{i\Theta_{\mu}},
\end{equation}

at the $m\geqslant 3$, there exist a class of exact solutions -
"color"  $U(\sigma{+}\mu)$-vector solitons with a nontrivial
interaction (intermode interchange). The conditions (4) imply that
every degree of freedom $\psi_n(1\leqslant n \leqslant
 m)$ in the system (3) has its vacuum
(condensate) of zero $\rho_n=0$ or finite $\rho_n\neq
 0$ density $\rho_n^2$ with an
asymptotic phase $\Theta_n$. In this case $m=\sigma+\mu$, and in
the rest $\sigma$ and $\mu$, take on the arbitrary values
${1,2,...,m}$.

\indent $N$-soliton formula of the $m$-component system (3-4)
depends definitely from the medium character $(\mathop{sign}
\ae=\pm 1)$ and in this sense it is a universal one for both
self-focusing (attractive,$+1$) and defocusing (repulsive, $-1$)
media of cube-nonlinearity.

\indent ISM in case of the system (3-4) faces the necessity of
analysis of the $(m+1)$-sheet Riemann surfaces and thereupon the
$N$-soliton solution is obtained by a more economical (in a
mathematical sense) method of Hirota [14]. It is shown that an
elastic (trivial) interaction of the mixed (bright and dark)
$U(1+1)$-vector solitons [9,12] is a consequence of the
$N$-soliton solution factorability for the system (3-4) in a
particular case of $m=2$. In the general position situation the
$N$-soliton solution is not factorizable and the interaction of
the color multisolitons has the inelastic nature (changing the
form at the energy conservation). There have been found special
cases when the interchange between nonlinear modes does not arise
and the $N$-soliton scaterring turns out to be a factorizable one.
\\
\\
\indent 1. Let's introduce the functions of Hirota

\begin{displaymath}
G_j=H\psi_j,~ G_j\in\mathbf{C}, H\in\mathbf{R},~ j=\overline{1,m}
\end{displaymath}

and make a transition $(\hat{L}_j\rightarrow\hat{D}_j)$ from the
linear operators -$\hat{L}$ to the bilinear $\hat{D}$ ones that
are defined hereinafter as
 \\
\begin{displaymath}
\hat{D}\ (U\cdot\ V)=(\hat{D}\ U)V-U(\hat{D}\ V).
\end{displaymath}

\indent Hereinafter variables $\zeta$ and $\xi$ at (1), (3-4) will
be attributed the sense of time $t$ and coordinate $x$.

\indent Taking into account the scaling changes $\vert \ae \vert
{=}2$, $c_j{=}c({>}0)$, $x{\rightarrow} x\sqrt{c}$, the system (1)
(due to the concept of Hirota) forms a bilinear family of the
compact $U(m)$-symmetry
\begin{equation}
\begin{array}{ll}
\hat{D}_1 G_{j}\cdot H=0,~~j=\overline{1,m}
\\
\hat{D}_2 H\cdot H=2 \delta \sum\limits_{k}^{m}\vert\
G_k\vert^{2},
\\
(\hat{D}_1=\textit{i}\hat{D}_t+\hat{D}_2,~\hat{D}_2=\hat{D}_x^2-\lambda).
\end{array}
\end{equation}

where $\lambda\in\mathbf{R}$ is an arbitrary parameter which will
be defined lower; $\delta = \mathop{sign} \ae$.

In the formalism of the bilinear operators for the functions $G_j$
and $H$ there is a representation in a series form due to the
formal parameter $\varepsilon$. Let's chose this representation in
such a way that it was coordinated with the non-trivial boundary
conditions (4), i.e.:


\begin{eqnarray}
G_j=\sum_{\nu=0}^{\infty}\varepsilon^{2\nu}
(g_{0\mu}g_{2\nu}^{\mu}\delta_{j\mu}+ \varepsilon
g_{0\sigma}g_{2\nu+1}^{\sigma}\delta_{j\sigma})
\nonumber\\
H=\sum_{\nu=0}^{\infty}\varepsilon^{2\nu} h_{2\nu};~~
g_0^{\mu}=h_0=g_{0\sigma}=1,\\
\delta_{\alpha\beta} - \textrm{Kronecker's symbol},~~~
j=\overline{1,m}\nonumber.
\end{eqnarray}

\indent It is obvious that functions $G_j$ determine $m$-component
field $(m{=}\sigma{+}\mu)$ in an arbitrary combination $\sigma$
and $\mu$ (for example, $\sigma$-bright solitons and $\mu$-dark
ones). We will get $N$-soliton solutions of the bilinear system
(5) following the standard scheme of Hirota ($R\thicksim G_j,
H;~j=\overline{1,m}$).


\begin{equation}
R=R_0\stackrel{\varepsilon^0}{\longrightarrow}\
R_1\stackrel{\varepsilon^1}{\longrightarrow}\ R_{2}
\stackrel{\varepsilon^2}{\longrightarrow} \ldots
\stackrel{\varepsilon^{N-1}}{\longrightarrow} R_N.
\end{equation}

Let's lay the vacuum solution
${g_{0\mu}=\rho_\mu\exp(\textit{i}\Theta_\mu)}$,
~$\Theta_\mu=k_\mu x-(k_\mu^2+\lambda)t$ in a zero order in
$\varepsilon$. One can determine
$\lambda=-2\delta\sum\limits_\mu^{m-\sigma}\rho_\mu^2$ from (5).
In the physics of optical solitons the sign function $
\delta=\mathop{sign}\ae$ defines a self-focusing ($\delta=+1$) or
defocusing ($\delta=-1$) character of the nonlinear medium. We
have the following in a zero order in $\varepsilon$:
{\setlength\arraycolsep{2pt}
\begin{eqnarray*}
g_1^{(j)}=\sum_{n=1}^N \gamma_n^{(j)}\exp(\eta_n),~
\eta_n=\zeta_nx+\textit{i}(\zeta_n^2+2\delta\sum_\mu^{m-\sigma}\rho_\mu^2)t,
\end{eqnarray*}}

where $\gamma_n^{(j)}$ and $\zeta_n$ are arbitrary complex
parameters. In a one-soliton ($N=1$) sector series (6) in the
scheme (7) stop in the second order in $\varepsilon$, in a
two-soliton ($N=2$) sector they stop in the fourth order in
$\varepsilon$ etc. The solutions of the system (5) describing the
propagation dynamics of $N$-solitons in two $\sigma$ and
$\mu$-sectors of $U(m)$-vector $\psi$-space (fixed on the 6 order
in $\varepsilon$) is represented as:

\begin{eqnarray}
\begin{array}{l}
H\psi_{\sigma}{=}\sum\limits_{n=1}^N
\hat{\eta}_n\{\varepsilon^{1}\gamma_n^{\sigma}{+}\sum\limits_{ij}^N
\hat{\eta}_i\hat{\eta}_j^{*} (\varepsilon^3 a_{nij}^{\sigma}
{+}{}\\
\qquad\qquad\qquad\qquad\qquad{}{+}\varepsilon^5\sum\limits_{l,m}^N
a_{nijlm}^{\sigma}\hat{\eta}_l\hat{\eta}_m^{*}{+}\ldots
\end{array}
\end{eqnarray}
\begin{equation}
\begin{array}{l}
H\psi_{\mu}{=}g_{0\mu}\{\varepsilon^{0}{+}\sum\limits_{ij}^N
\hat{\eta}_i\hat{\eta}_j^{*} [\varepsilon^2
a_{ij}^{\mu}{+}\sum\limits_{lm}^N \hat{\eta}_l\hat{\eta}_m^{*}
(\varepsilon^4 a_{ijlm}^{\mu}{+}{}\\
\qquad\qquad\qquad\qquad\qquad{}{+}\varepsilon^6
\sum\limits_{q,r}^N
a_{ijlmqr}^{\mu}\hat{\eta}_q\hat{\eta}_r^{*}{+}\ldots
\end{array}
\end{equation}
\begin{eqnarray*}
\begin{array}{l}
H{=}\varepsilon^{0}{+}\sum\limits_{ij}^N
\hat{\eta}_i\hat{\eta}_j^{*} [\varepsilon^2 a_{ij}{+}
\sum\limits_{lm}^N \hat{\eta}_l\hat{\eta}_m^{*}
(\varepsilon^4 a_{ijlm}{+}{}\nonumber\\
\qquad\qquad\qquad\qquad\qquad{}{+}\varepsilon^6
\sum\limits_{q,r}^N
a_{ijlmqr}\hat{\eta}_q\hat{\eta}_r^{*}{+}\ldots\nonumber
\end{array}
\end{eqnarray*}

Here $\varepsilon=1~ \textrm{(R.Hirota)},~ \hat{\eta}_n=
\exp(\eta_n)$;
\begin{eqnarray}
a_{nij}^{\sigma}=\frac{\zeta_{ni}^{-}}{\zeta_{nj}\zeta_{ij}}(\gamma_n^{\sigma}
\tilde{a}_{ij}-\gamma_i^{\sigma} \tilde{a}_{nj}), \nonumber\\
a_{ijlm}=\frac{\zeta_{il}^{-}\bar{\zeta}_{jm}^{-}}{\zeta_{ij}\zeta_{im}\zeta_{lj}\zeta_{lm}}
(\tilde{a}_{ij}\tilde{a}_{lm}-\tilde{a}_{im}\tilde{a}_{lj}),\nonumber\\
a_{nijlm}^{\sigma}=\frac{\zeta_{ni}^{-}\zeta_{nl}^{-}}{\zeta_{nj}\zeta_{nm}}
\gamma_n^{\sigma}a_{ijlm}+\left\{\begin{array}{ll}\scriptstyle
n\leftrightarrow
i\leftrightarrow l\nonumber\\
\scriptstyle~~j\leftrightarrow m
\end{array} \right\},\nonumber\\
a_{ijlmqr}=\frac{\zeta_{qi}^{-}\zeta_{ql}^{-}\bar{\zeta}_{rj}^{-}\bar{\zeta}_{rm}^{-}}
{\zeta_{qj}\zeta_{qm}\zeta_{ir}\zeta_{lr}} a_{ijlm}
a_{qr}+\left\{\begin{array}{cc}\scriptstyle q\leftrightarrow
i\leftrightarrow l\nonumber\\
\scriptstyle r\leftrightarrow j\leftrightarrow m
\end{array} \right\},\nonumber\\
a_{\ldots ij \ldots}^{\mu}=z_{ij}^{\mu}a_{\ldots ij \ldots},
z_{ij}^{\mu}=-\frac{z_{i\mu}}{\bar{z}_{j\mu}},
z_{j\mu}=\zeta_j-ik_{\mu},\\
\tilde{a}_{ij}=\frac{\sum\limits_{\sigma=1}^{m-\mu}\gamma_i^{\sigma}\bar{\gamma}_j^{\sigma}}{\zeta_{ij}(\delta+\sum\limits_{\mu}^{m-\sigma}\rho_{\mu}^2/z_{i\mu}\bar{z}_{j\mu})}
, ~a_{ij}=\frac{\tilde{a}_{ij}}{\zeta_{ij}},
\nonumber\\
\zeta_{lm}=\zeta_{l}+\bar{\zeta}_{m},~\zeta_{lm}^{-}=\zeta_{l}-\zeta_{m},~~\bar{O}\equiv
O^{*}, k_{\mu}\in\mathbf{R}.\nonumber
\end{eqnarray}

\indent It is quite evident from here that the two-soliton $(N=2)$
solution stops at the 4th order in $\varepsilon$. At the same time
one can assure that the formulas (8-9) correspond to the exact
three-soliton $(N=3)$ solution of the system under discussion (5).
The solutions of the higher order are not given here because of
their awkwardness.
\\
\\
\indent 2. One-soliton $(N{=}1)$ solution of
Shr\"{o}dinger's mixed $U(\sigma{+}\mu)$-vector
nonlinear model (3-4) from (8-10) takes the form:


\begin{eqnarray}
\left( \begin{array}{c}
\psi_{\{\sigma\}}\\
\psi_{\{\mu\}}
    \end{array}\right)
    =H^{-1}\left(
    \begin{array}{c}
\gamma_1^{\sigma} e^{\eta_1}\\
g_{0\mu}(1+a_{11}^{\mu}e^{\eta_1+\bar{\eta}_1})\\
    \end{array}\right)
\end{eqnarray}
where Hirota function is $H=1+a_{11}e^{\eta_{1}+\bar{\eta}_{1}},$
\begin{displaymath}\begin{array}{c}
a_{11}^{-1}=\zeta_{11}^2(\delta+\sum\limits_{\mu}^{m-\sigma}\rho_{\mu}^2\vert
z_{1\mu}\vert^{-2})
/\sum\limits_{\sigma}^{m-\mu}\vert\gamma_1^{\sigma}\vert^2,\\
~a_{11}^{\mu}=z_{11}^{\mu}a_{11},~
z_{11}^{\mu}=-\exp(2i\phi_{1\mu}),\\
~\phi_{1\mu}=\arctan((\mathbf{Im}\zeta_1-k_{\mu})/\mathbf{Re}\zeta_1).
\end{array}
\end{displaymath}
\indent

As one can see the $U(\sigma{+}\mu)$-vector soliton of the mixed
color (11) is a dynamics-and-topological formation and in
particular cases coincides with known earlier one-color
bright-vector $(\psi_{\{\mu\}}{=}0,~\delta{=}{+}1)$ [8] and
dark-vector $(\psi_{\{\sigma\}}{=}0,~\delta{=}{-}1)$ [11,12]
solitons. However, there is a principally new thing, i.e. the fact
what the exact solution (11), unlike the vector solitons [8,11,12]
has its place in the system (3) both in the cases of attraction
(self-focusing, $\delta{=}{+}1$) and repulsion (defocusing,
$\delta{=}{-}1$). Besides that, one should mention that a
universal color $U(\sigma{+}\mu)$-vector soliton (11) can be in
several states determined by its dynamics-and-topological nature.
It is convenient to interpret these states in the language of
particles. Let's denote the admissible isotopic states of the
color $U(\sigma{+}\mu)$-vector soliton by the symbol
$\{\sigma,\mu\}$, where $\sigma{+}\mu{=}m$. Then by analogy with a
quantum chromodynamics, the state with a mixed color
$\{\sigma{\neq}0,~\mu{\neq}0\}$ may be supposed to be "aromatic"
and the one with a mixing miss $(\{0,\mu\}, \{\sigma,0\})$ - "no
aromatic" (one-color). In the given analogy the $U(m)$-vector
soliton of a mixed color has the intrinsic structure and the
existence of different states is natural for such a compound
particle. For example, in case of Shr\"{o}dinger's $U(5)$-vector
nonlinear model the solution (11) for one $\{3,2\}$ of the 4
admissible ($\{1,4\},\{2,3\},\{3,2\},\{4,1\}$) aromatic states of
the color $U(3{+}2)$-vector solitons has the form:


\begin{eqnarray}
\left( \begin{array}{c}
\psi_{1}\\
\psi_{2}\\
(\psi_{3}\psi_{4}\psi_{5})^t
    \end{array}\right)
   = \left(
    \begin{array}{c}
A_1(\tanh X+i\tan\phi_1)e^{i\Theta_1}\\
A_2(\tanh X+i\tan\phi_2)e^{i\Theta_2}\\
(B_3B_4B_5)^tsech Xe^{i\Theta}
    \end{array}\right).
\end{eqnarray}

\begin{displaymath}\begin{array}{c}
A_{\mu}=\rho_{\mu}\cos\phi_{\mu},
\Theta_{\mu}=k_{\mu}x-(k_{\mu}^2-2\delta\sum\limits_{\mu=1}^2\rho_{\mu}^2)t+\phi_{\mu},\\
\phi_{\mu}=\arctan((v-2k_{\mu})/u);~~2X=u(x-vt-x_0),\\
B_{\sigma}=\gamma^{\sigma}[(\sum\limits_{\mu=1}^2A_{\mu}^2+\delta
u^2/4)/
\sum\limits_{\sigma=3}^5\vert\gamma^\sigma\vert^2]^{\frac{1}{2}},\\
2\Theta=vx+(u^2-v^2+8\delta\sum\limits_{\mu=1}^2\rho_{\mu}^2)t/2;\\
\mu=1,2;~~\sigma=3,4,5;
\end{array}
\end{displaymath}

where $u{=}2\mathbf{Re}\zeta_1$ and $~v{=}2\mathbf{Im}\zeta_1$ are
soliton's reverse width and velocity. It is clear that a mixed
$U(5)$-vector soliton (12) consists of two dark $(\psi_1,\psi_2)$
and three bright $(\psi_3,\psi_4,\psi_5)$ components (nonlinear
modes). It is also obvious that the number of all the
$U(5)$-vector soliton admissible states is equal to 6, however two
of them are no aromatic (one-color) vector solitons: a
bright-vector one $\{5,0\}$ in case of self-focusing medium
$(\delta=+1)$ and a dark-vector one $\{0,5\}$ in case of
defocusing medium $(\delta=-1)$. \indent The change of the medium
refraction index $\bigtriangleup
n^2\sim\vert\psi_1\vert^2+\vert\psi_2\vert^2+\ldots+\vert\psi_5\vert^2$
induced by the $\psi_1,\ldots, \psi_5$ components interaction may
be calculated by the forward substitution of an explicit solution
(12). However, there comes a universal formula from the bilinear
system of equations (5) for the whole family of Shr\"{o}dinger's
$U(m)$-vector nonlinear models,
\begin{equation}
\bigtriangleup
n^2=\sum_{\mu}\rho_{\mu}^2+\delta\frac{d^2}{dx^2}\ln H,
\end{equation}
 which permits to determine the value of $\bigtriangleup n^2$ of
 the Hirota unified \mbox{function $Н$.} So, for example, in case of
 consideration of the above $U(5)$-model $\mu=\overline{1,2}$,
 $H=1+a_{11}e^{\eta_1+\bar{\eta}_{1}}$. There will be derived from
 the (13)
\begin{displaymath}
\bigtriangleup
n^2=\rho_{1}^2+\rho_{2}^2\pm\left(\frac{u^2}{4}\right)sech^2\left[\frac{u(x-vt-x_0)}{2}\right]
\end{displaymath}

for the self (de) focusing $(+(-))$ medium.

\indent One should mention that presence of vacuum-condensate  in
the system with a finite density $\rho_{\mu}^2$ in the defocusing
$(\delta{=}{-}1)$ media imposes a natural limitation on the color
vector soliton characteristics:
$u^2\leqslant4\sum\limits_{\mu}\rho_{\mu}^2\cos^2\phi_{\mu}$.
Nevertheless as far as the number of the admissible aromatic
states of the color $U(m)$-vector soliton is equal to $(m-1)$, the
detection of exactly the same states in the multimode optical
systems may turn out an event, which is more probable than
one-color states whose number is equal to two.
\\
\\
\indent 3. Two-soliton ($N=2$) solution and dynamics
of the mixed color multisoliton interaction.

\indent Let's show that the interaction of the color multisolitons
(8,9) in Shr\"{o}dinger's mixed $U(m)$-vector nonlinear model
(3,4) is nontrivial (changing a form without any energy loss) at
$m\geqslant 3$ and there has a place the intermode interchange
(energy) that is proportional to the nonlinear modes intensity of
solitons. Let's study for this goal (without community limitation)
asymptotic $(t\rightarrow\pm\infty)$ behavior of the color
multisolitons (8,9) at $N=2$.

\indent Two-soliton $(N=2)$ solution from (8,9) takes the form
$(\sigma{+}\mu{=}m)$:
\begin{eqnarray}
\left(\!\!\!\begin{array}{l} \scriptstyle{\psi_{\{\sigma\}}}
\\
\scriptstyle{\psi_{\{\mu\}}}
    \end{array}\!\!\!\right)
    \!\scriptstyle{{=}H^{-1}}\!\left(\!\!\!\begin{array}{l}
\scriptstyle{\gamma_1^{\sigma} e^{\eta_1}{+}\gamma_2^{\sigma}
e^{\eta_2}{+}\sum\limits_j^2a_{12j}^{\sigma}e^{\eta_1{+}\eta_2{+}\bar{\eta}_j}}
\\
\scriptstyle{g_{0\mu}(1{+}\sum\limits_{i,j}^2a_{ij}^{\mu}e^{\eta_i{+}\bar{\eta}_j}
{+}a_{1122}^{\mu}e^{\eta_1{+}\bar{\eta}_1{+}\eta_2{+}\bar{\eta}_2})}
    \end{array}\!\!\!\right),
\end{eqnarray}
where Hirota function
\begin{displaymath}
H=(1+\sum_{i,j}^2a_{ij}e^{\eta_i+\bar{\eta}_j}+a_{1122}e^{\eta_1+\bar{\eta}_1+\eta_2+\bar{\eta}_2})
\end{displaymath}

Let $v_1>v_2(\mathbf{Im}\zeta_1{>}\mathbf{Im}\zeta_2)$, where
$v_n$ is $S_j^n$-soliton velocity in $j$ mode
$(j{=}1,2,\ldots,\mu,\mu{+}1,\ldots,\mu{+}\sigma)$. Solution (14)
at $t\rightarrow\pm\infty$ on the paths $\xi_n=x-v_nt$ of separate
solitons fall into the sum of the free one-soliton solution of the
type (12):
\begin{equation}
\psi_{\{j\}}(x,t)\bigg\vert_{t\rightarrow \pm\infty}{=}\sum_n
C_j^{n\pm}S_{\{j\}}^n(x{-}v_nt,x_{0n}^{\pm})e^{i\Theta_{nj}}
\end{equation}
Here $C_j^{n\pm}$-amplitude, $S_{\{j\}}^n$-envelope of $j$ mode of
$n$ soliton $n=1,2;~j=\sigma,\mu$:
\begin{displaymath}
\begin{array}{c}
S_{\{\sigma\}}^n=sech Y_n^{\pm},~ S_{\{\mu\}}^n=\tanh
Y_n^{\pm}+i\tan\phi_{n\mu},\\
Y_n^{\pm}=u_n(x-v_nt-x_{0n}^{\pm})/2,\\~
\phi_{n\mu}{=}\arctan((v_n-2k_{\mu})/u_n),~
u_n{=}2\mathbf{Re}\zeta_n,~v_n{=}2\mathbf{Im}\zeta_n.
\end{array}
\end{displaymath}

Amplitudes $C_j^{n\pm}$ of $S_j^n$ solitons before (--) and after
(+) the interaction are connected by the relations
$C_j^{n+}=\hat{S}_j^nC_j^{n-}$, where $\hat{S}$-special matrix,
converting asymptotic at $t\rightarrow -\infty$ in the asymptotic
at $t\rightarrow +\infty$:
\begin{eqnarray}
\begin{array}{c}
\hat{S}_{\sigma}^1=\tilde{\zeta}_{12}(1-s_1\gamma_{21}^\sigma)(1-s_1s_2)^{-\frac{1}{2}},\\
~2C_{\sigma}^{1-}=\gamma_1^{\sigma}(a_{11})^{-\frac{1}{2}},
\hat{S}_{\mu}^1=e^{i(2\phi_{2\mu}-\pi)},\\
~C_{\mu}^{1-}=\rho_{\mu}\cos\phi_{1\mu}e^{i(\phi_{1\mu}+\pi)},\\
\hat{S}_{\sigma}^2=\tilde{\zeta}_{21}(1-s_2\gamma_{12}^\sigma)^{-1}(1-s_1s_2)^{\frac{1}{2}},\\
~2C_{\sigma}^{2-}=a_{121}^{\sigma}(a_{1122}a_{11})^{-\frac{1}{2}},
\hat{S}_{\mu}^2=e^{-i(2\phi_{1\mu}-\pi)},\\
~C_{\mu}^{2-}=\rho_{\mu}\cos\phi_{2\mu}e^{i(2\phi_{1\mu}+\phi_{2\mu})};\\
\gamma_{ij}^{\sigma}=\frac{\gamma_{i}^{\sigma}}{\gamma_{j}^{\sigma}},
~s_1=\frac{\tilde{a}_{12}}{\tilde{a}_{22}},~
s_2=\frac{\tilde{a}_{21}}{\tilde{a}_{11}},
~\vert\tilde{\zeta}_{12}\vert=\vert\tilde{\zeta}_{21}\vert=1.
\end{array}
\end{eqnarray}
Hence it is quite clear that solitons velocities $v_n$-motion
invariants, and phases $X_{0n}^{\pm}$ and amplitudes $C_j^{n\pm}$
are not such. As $\vert\hat{S}_{\sigma}^n\vert\neq 1$ and
$\vert\hat{S}_{\mu}^n\vert=1$ in the situation of a general
position, it is quite obvious that the interaction of the mixed
color $U(\sigma{+}\mu)$-vector solitons (14) has a nontrivial
(inelastic) character. As a result of such an interaction between
nonlinear modes there arises the intensity interchange
$\sim\vert\hat{S}_{j}^n\vert^2$. The intermode exchange initiates
energy redistribution in the components (nonlinear modes) of the
color vector solitons. However, the given interchange phenomena in
the $\sigma$- and $\mu$- modes have their peculiarities:
$\sigma$-modes interchange by the finite energy
$\sim\vert\hat{S}_{\sigma}^n\vert^2$ and maintain the sign; $\mu$-
modes maintain energy $(\vert\hat{S}_{\mu}^n\vert^2=1)$, but
change their polarity and attain an additional phase jump
$\sim(2\phi_{n\mu}+\pi)$ as an interaction result. It is clear
that $\mu$- modes interact only with different values of phases.

\indent Asymptotic analysis shows in general that the exchange
between the components of a separate color soliton is not
arbitrary (chaotic) but is correlated with the commensurable
changes in the components of all the other solitons. A nontrivial
interaction between the color multisolitons (14) and the
admissible scenarios of the intermode exchange in the system (3,4)
are regulated by the laws of conservation:
$a)~\sum\limits_j^m\vert C_j^{n-}\vert^2=\sum\limits_j^m\vert
C_j^{n+}\vert^2$ -total intensity of a separate soliton $S_j^n$
and $b)~\sum\limits_n(\sum\limits_j\vert
C_j^{n-}\vert^2)=\sum\limits_n(\sum\limits_j\vert
C_j^{n+}\vert^2)$ - full intensity of the all the solitons before
(--) and after (+) interaction. Validity of the given laws may be
easily seen from the asymptotic formulas (16). Besides this,
shifts of their inertia centers $\triangle
X_n=X_{0n}^{+}-X_{0n}^{-},~\triangle
X_n=(-1)^{n+1}2\zeta_{nn}^{-1}\ln \chi$, where

\begin{eqnarray}
\begin{array}{c}
\chi=\vert\frac{\zeta_1-\zeta_2}{\zeta_{12}}\vert^2
\sqrt{1+\frac{\zeta_{11}\zeta_{22}}{\vert\zeta_1-\zeta_2\vert^2}(1-\epsilon)},\\
\\
\epsilon=\frac{(\delta+\sum\limits_{\mu}^{m-\sigma}\rho_{\mu}^2/\vert
z_{1\mu}\vert^2)
(\delta+\sum\limits_{\nu}^{m-\sigma}\rho_{\nu}^2/\vert
z_{2\nu}\vert^2)\sum\limits_{\sigma,\tau}^{m-\mu}
\gamma_1^{\sigma}\gamma_2^{\tau}\bar{\gamma}_1^{\tau}\bar{\gamma}_2^{\sigma}}
{\vert\delta+\sum\limits_{\mu}^{m-\sigma}\rho_{\mu}^2/
z_{1\mu}\bar{z}_{2\mu}\vert^2\sum\limits_{\sigma,\tau}^{m-\mu}
\vert\gamma_1^{\sigma}\vert^2\vert\gamma_2^{\tau}\vert^2}
\end{array}
\end{eqnarray}

arising as a result of solitons interaction obey the condition of
Sudzuki-Zakharov-Shabat (the law of solitons center of inertia
conservation): \\ \mbox{$\zeta_{11}\triangle X_1+\zeta_{22}\triangle
X_2=0$.} The last one is a consequence of the value conservation
$I_{tot}=\int\sum\limits_n\vert\psi_n\vert^2dx$ in time $t$. The
laws of conservation and strict formulas that are mentioned above
permit to determine the interchange "kinetics" and admissible
scenarios of the intermode switches in the system of the color
multisolitons (14) by the quantitative fashion. However, let's pay
attention to the factor $\epsilon$ in (17).

\indent Due to the apparent many particle effects in $\epsilon$
the shifts of the inertia centers of the color solitons $\triangle
X_n$ don't accord with a conventional feature of factorization
that is traditional for the ordinary solitons. Thus, $\epsilon$
(in a concentrated form) points at a complex nature of interaction
of the mixed $U(\sigma+\mu)$-vector solitons (14). In the
situation of a general position (solitons parameters $\gamma_n^j,
\zeta_n, (z_{j\mu})$, vacuum densities $\rho_{\mu}^2$ and
component number $\sigma{+}\mu{=}m$ - arbitrary), the shifts
$\triangle X_n$ of solitons because of $\epsilon$ can't be
presented in a two-particle form, $N$-soliton scattering doesn't
come to the pair one and the interaction of mixed
$U(\sigma{+}\mu)$-vector solitons at $\sigma{\geqslant}2$ is
nontrivial (changing the form at the energy conservation). In a
particular case of the linear dependence of the parameters
$\gamma_i^{\sigma}\gamma_j^{\nu}{-}\gamma_i^{\nu}\gamma_j^{\sigma}{=}0$
their influence on $\epsilon$ disappear, the vacuum contribution
are balanced and the center shifts $\triangle X_n$ admit
two-particle representation. Consequently, the $N$-soliton
solution is factorized and the soliton interaction becomes elastic
($\vert\hat{S}_j^n\vert{=}1,~j{=}\sigma,\mu$). Besides this, from
the (17) directly results  that in a special case of Manakov's
mixed $U(2)$-model [12] $\sigma{=}\mu{=}1$ the interaction of the
mixed $U(1{+}1)$-vector solitons is strictly elastic. In all the
other cases at $m{\geqslant}3$ color $U(\sigma{+}\mu)$-vector
solitons $(m{=}\sigma{+}\mu)$ interact by the nontrivial fashion
and there is an energetic interchange between their nonlinear
modes that is consistent with the above-mentioned laws of
conservation.

\indent In conclusion the authors express their gratitude to
V.E.Zakharov for his attention to their work, S.V.Manakov and V.
G.Marikhin  for their useful remarks. The work was partially
fulfilled at Landau ITP.

\end{document}